# Strain-balanced type-II superlattices for efficient multi-junction solar cells

A. Gonzalo (1), A. D. Utrilla (1), D. F. Reyes (2), V. Braza (2), J. M. Llorens (3), D. Fuertes Marrón (4), B. Alén (3), T. Ben (2), D. González (2), A. Guzman (1), A. Hierro (1), J. M. Ulloa (1)*

(1) *Institute for Systems based on Optoelectronics and Microtechnology (ISOM), Universidad Politécnica de Madrid, Avda. Complutense 30, 28040 Madrid, Spain*

(2) *Departamento de Ciencia de los Materiales e IM y QI, Universidad de Cádiz, 11510 Puerto Real (Cádiz), Spain*

(3) *IMM-Instituto de Microelectrónica de Madrid (CNM-CSIC), Isaac Newton 8, PTM, E-28760 Tres Cantos, Madrid, Spain*

(4) *Instituto de Energía Solar (IES), Universidad Politécnica de Madrid, Avda. Complutense 30, 28040 Madrid, Spain*

* electronic mail: jmulloa@isom.upm.es

**Abstract**

We propose type-II GaAsSb/GaAsN superlattices (SLs) lattice-matched to GaAs as a novel material for the 1 eV sub-cells present in highly efficient GaAs/Ge-based multi-junction solar cells. We demonstrate that, among other benefits, the spatial separation of Sb and N allows a better control over composition and lattice matching, avoiding the growth problems related to the concomitant presence of both elements in GaAsSbN layers. This approach not only reduces clustering and improves crystal quality and interface abruptness, but also allows for additional control of the effective bandgap in the 1.0-1.15 eV spectral region through the SL period thickness. The optimized SL structure exhibits a type-II band alignment and strong electronic coupling at 0 V. Both effects cooperate to increase the minority carrier collection and leads to a net strong enhancement of the external quantum efficiency (EQE) under photovoltaic conditions with respect to bulk layers of equivalent thickness.

In the solar cell community, there is a quest for surpassing the 50 % efficiency psychological barrier, which implies that half of the light energy coming from the Sun would be transformed into electric power. The solar cell efficiency tables[1] show how close we are to reach that limit: 46.0 % under concentration and 38.8 % under AM1.5G conditions. Theoretical efficiency limits make use of optimum multi-junction designs with the ideal lattice constant-bandgap energy combination but, sometimes, lack of suitable, easy to control semiconductor materials hinders their implementation. For instance, in the standard (Al)InGaP/(In)GaAs/Ge solar cell structure, the addition of a lattice-matched sub-cell tuned to the 1.0-1.15 eV spectral range provides such an optimum multi-layer design.[2,3] Under standard AM1.5G conditions, detailed-balance calculations predict 44.4 % efficiency for a 3-junction AlInGaP(1.9 eV)/GaAs(1.4 eV)/1.0 eV solar cell and 47.7 % for a 4-junction AlInGaP(1.9 eV)/GaAs(1.4 eV)/1.0 eV/Ge(0.66 eV) solar cell. Both designs would easily leave behind the 50 % limit when operated under concentration.[4] After the two consecutive world efficiency records set by Solar Junction CA,[5] GaInNAs in combination with Sb has gained a great interest as such 1.0 - 1.15 eV material. The fabricated GaInP/GaAs/GaInAsNSb solar cells had an efficiency of 44.0 % under concentration, below the theoretical limit, and was limited by epitaxial growth problems which are intrinsic to quaternary and quinary materials. These problems seriously affect carrier dynamics and are likely the reason for the lack of success in achieving ultimate solar cell performance beyond the 50 % barrier.[6] In this work, we show how the use of GaAsSb/GaAsN short period SLs with type-II band alignment might solve these problems at once.

Monolithic multi-junction engineering confronts an issue of paramount importance: the different sub-cells have to maintain a very low stress level during epitaxial growth, avoiding the introduction of misfit dislocations, which would degrade the device performance.[7,8] Therefore, materials lattice-matched to GaAs/Ge and with a 1.0 eV or 1.15 eV bandgap are being extensively investigated. Among them, dilute nitride semiconductor alloys, such as GaInNAs or GaAsSbN, can fulfill both requirements and be used for this purpose.[9] In these highly mismatched alloys, the addition of small amounts of N leads to a dramatic reduction of the bandgap, which is explained in the framework of the band anti-crossing (BAC) model as the result of a strong interaction between the localized N states and the conduction band of the matrix.[10] However, GaAsSbN alloys show important potential advantages over GaInNAs, because the surfactant effect of Sb atoms facilitates the two-dimensional growth of the material.[11] It also has the unique property of allowing an independent tuning of both conduction and valence band energies by controlling the N and Sb contents, respectively.[12] Besides, GaAsSbN can remain lattice-matched to GaAs if the condition $[Sb] \approx 2.8 \times [N]$ is fulfilled,[13] since Sb compensates the tensile strain induced by N. The GaAsSbN alloy has already been applied to solar cell technology, both as a thick layer[14-17] and as a capping layer over InAs quantum dots.[18,19] Nevertheless, the obtained solar cell performance is not satisfactory up to now. This is due to the fact that GaAsSbN faces important epitaxial growth problems such as phase separation (because of its large miscibility gap), clustering, difficult composition control (3 group-V atoms competing for the same lattice position), or N-related point defects and localized electronic states (as in any other dilute nitride alloy).[20-22]

The only way to overcome these problems in the growth of the quaternary alloy is to resort to new growth approaches. In this work, we show that an effective way to obtain a material of extremely high crystal quality and fully stoichiometric is by splitting the quaternary into two ternaries and proceed by growing a SL structure; the spatial separation of N and Sb dramatically reduces miscibility problems. The SL structures can be fabricated with type-I or type-II band alignment (electrons confined in GaAsN and holes in GaAsSb) offering additional advantages over the bulk counterparts, such as long carrier lifetimes (and, therefore, enhanced collection efficiency)[23,24] and effective bandgap tunability through period thickness. Moreover, independent conduction and valence band edge manipulation makes it possible to reproduce nominal bandgap energies of the quaternary alloy using only half the amount of Sb and N. However, the use of SLs might also bring a major drawback, which is the increase of the series resistance and thus, a severe reduction of the quantum efficiency, as many of the photogenerated carriers could not reach the contacts. Our results show that this is not the case in the type-II GaAsN/GaAsSb system when the thickness of the SL period is reduced to 6 nm. For such thin stacks, strong quantum tunneling comes into play producing devices with the same quantum efficiency at zero and reverse biasing, i.e. achieving a complete carrier collection under photovoltaic conditions.

## 2. Results and Discussion

### 2.1. Interface Sharpness and Crystallinity

In the first series of samples, different SL structures containing 18 periods with a constant period thickness of 12 nm (6 nm+6 nm) as active layer are compared. The total thickness of the SL is 200 nm and they consist of: GaAsSb/GaAs (Sample *SL-Sb*), GaAs/GaAsN (sample *SL-N*), GaAsSbN/GaAs (sample *SL-I* with expected type-I confinement) and GaAsSb/GaAsN (sample *SL-II* with expected type-II confinement). A fifth sample with a 200 nm-thick GaAsSbN bulk layer (sample *bulk*) was grown as a reference to complete the series. A scheme of the epitaxial structure and the expected band structure of samples *SL-I* and *SL-II* are shown in **Figure 1**a,b respectively (see Figure S1a,b,c, Supporting Information for the sketch of rest of the samples). All the samples were grown under the same Sb and N nominal fluxes; therefore, only half the amount of N and Sb was nominally used to fabricate the SL structures as compared to the bulk, and the amount of low bandgap active material was half in these structures.

The samples were structurally analyzed by transmission electron microscopy (TEM). Dark field 002 representative images of samples *SL-I* and *SL-II* are shown in Figure 1c,d (see Figure S1d,e,f, Supporting Information for the rest of the samples). In this chemically sensitive TEM imaging mode, Sb-rich regions appear brighter whereas N-rich regions appear darker than GaAs regions. It can be observed that both samples exhibit flat growth fronts and there is no evidence of dislocations or any other sort of extended defects. A detailed analysis demonstrates that *SL-I* and *SL-II* are completely pseudomorphic, and that all layers have similar compositions along the structure. The periodicity is regular throughout the whole structure with estimated period thicknesses of 12.7 nm for *SL-I* and 12.9 nm for *SL-II*, slightly larger than the nominal value of 12 nm due to a minor increase in the growth rate. Also, the interface contrast appears more abrupt in *SL-II*, indicative of reduced Sb segregation.[25]

All samples from this set were investigated by high-resolution x-ray diffraction (HR-XRD). **Figure 2** shows the corresponding omega-2 theta scans around the (004) GaAs reflection. The HR-XRD spectra of the ternary SLs *SL-Sb* and *SL-N* allow estimating the Sb and N composition to be 3.25 % and 1.20 % in the respective layers, considering a completely pseudomorphic epitaxy as observed by TEM. These two values fulfill the lattice matching condition for GaAsSbN on GaAs, as it is evidenced by the symmetrical position of the main SL peaks with respect to the substrate peak. We also observe that the secondary peaks are considerably less intense in *SL-Sb* than in *SL-N*, meaning that the *SL-Sb* sample has worse periodicity, likely due to Sb segregation in this type of structures. Assuming that there is no Sb-N interaction during growth that could modify the composition, lattice matching is expected from the nominal growth parameters also in the rest of the samples. However, in the *SL-I* and *bulk* HR-XRD spectra, the main peak appears shifted towards the tensile region of the spectrum. This indicates that the composition is altered from that of the ternaries due to the concomitant presence of N and Sb in the growth front. This effect is particularly strong in the bulk sample.

The main motivation in this work is to explore the alternate introduction of Sb and N atoms in the growth surface to avoid the interaction between both species. We expect that the composition in the pseudo-quaternary structure can consequently be accurately controlled. Figure 2 demonstrates that this is the case for sample *SL-II* where we find a perfectly lattice-matched spectrum implying that the expected contents were incorporated and the overall strain precisely compensated. Moreover, the narrower main peak in the SLs as compared to the bulk structure indicates a better strain and composition homogeneity, which means that clustering effects are significantly reduced in the SL structures. Remarkably, the secondary peaks are also narrower and more intense in *SL-II* as compared to *SL-I* reflecting the higher interface quality when Sb and N are incorporated separately, in agreement with what was observed by TEM.

**2.2. Energy and Radiative Lifetime Tuning**

The higher control over composition achievable with the type-II SL as compared to the type-I SL or the bulk is an important feature for energy tuning of the bandgap in multi-junction solar cell applications. **Figure 3**a shows photoluminescence (PL) spectra measured in the samples described above at 15 K. The shift of the *SL-II* emission peak energy with respect to the bandgap of GaAs (dashed vertical line) is 280 meV. This fits almost perfectly with the combined energy shifts observed for the ternary samples *SL-Sb* (27 meV) and *SL-N* (250 meV). This is a quite unique property of this system which suggests that tuning of the valence and conduction bands, and thus the bandgap energy value, can be done setting independently the Sb and N concentrations. Such independent control is hard to reproduce in the bulk quaternary alloy. All samples were fabricated using the same nominal ternary concentrations and, yet, the sample *bulk* significantly redshifts from the nominal bandgap value. Finite element calculations shown below suggest that quantum confinement size effects in the 12 nm period SLs can only partially explain this difference (~39 meV). The redshift of the PL peak energy and the tensile position of the HR-XRD central peak shown in Figure 2 rather points to an unwanted incorporation of N in this sample (and, to a lesser extent, also in the *SL-I* sample), which was suppressed in the *SL-II* sample using alternate deposition of the ternary compounds.

Time resolved-photoluminescence (TR-PL) decay curves measured at the PL peak energy provide further details about the carrier confinement within each sample. As shown in Figure 3b, *SL-I* and *bulk* samples have similar decay dynamics, clearly different from those of sample *SL-II*. A multi-exponential fitting analysis is necessary to describe the decay dynamics across the full-time range (Table S1, Supporting Information). Yet, the larger differences occur for long times after the excitation where the decay of the luminescence of sample *SL-II* becomes three times slower (decay constant changes from ~16 to ~49 ns). The existence of this significantly longer radiative lifetime at the PL peak energy only for sample *SL-II* strongly supports the predominance of type-II band alignment and recombination. Slower carrier recombination shall lead to an enhanced carrier extraction and, therefore, an improved photocurrent. Remarkably, in this sample, not only the PL band is narrower but the integrated PL emission is the most intense despite the longer carrier lifetime (see Figure 3a).[26] This is a clear indication of the improved crystal and interface quality produced by our method, also underlined by the fact that the PL of bulk GaAsSbN layer, despite having twice as much active material, is much weaker.

In order to investigate the formation of minibands and the effective bandgap tunability through the SL period thickness, four additional samples were grown: 200 nm-thick type-II SLs with the same N and Sb contents as the previous set of samples, but different period thickness: 3 nm, 6 nm, 12 nm and 20 nm. HR-XRD measurements indicate that all the samples are lattice-matched to GaAs and that the period thicknesses agree with the nominal values: 3.1 nm, 6.4 nm, 12.6 nm and 19.1 nm, respectively. Tuning in relevant spectral regions shall be demonstrated at room temperature. Thus, the effective bandgap energies from these SLs have been obtained from photoreflectance (PR) measurements at room temperature, shown in **Figure 4**a. The PR spectra have been analyzed through the third derivative functional form (TDFF) method.[27] The lowest energy critical point obtained thereof is associated with the optical transition between electron and hole ground states. As shown in Figure 4b, the effective bandgap energy increases rapidly as the SL period decreases, indicative of a quantum confinement size effect. The bandgap energy tuning capability saturates for periods larger than 20 nm. Noticeably, a bandgap tunability of more than 100 meV is achieved within the investigated range. These experimental values have been compared with electronic band structure calculations; the calculated ground state energies shown in Figure 4b follow the same trend of the PR results, with a rigid shift to lower energies likely due to the simplified approximations assumed in the model. These findings confirm the expected quantum confinement in the structures and the ability to tune the effective bandgap at room temperature by varying the period thickness.

### 2.3. Single-Junction Solar Cell Performance

Efficient carrier transport through the epitaxial structure is mandatory to assess the performance of these SL structures as solar sub-cells. We have investigated the EQE at room temperature of five p-i-n diodes with different 750 nm-thick active regions: one GaAsSbN/GaAs type-I SL structure with 12 nm period (*SL-I$_{12}$*), two GaAsSb/GaAsN type-II SL structures with 12 nm (*SL-II$_{12}$*) and 6 nm (*SL-II$_6$*) period, a bulk GaAsSbN layer (*bulk*) and a GaAs reference sample (*GaAs*) for comparison. The Sb and N-containing samples were all grown under the same nominal fluxes. Comparing the EQE of devices based on low-dimensional structures and bulk layers with the same effective bandgap is not straightforward. On one hand, the total intrinsic region thickness should be the same to do the comparison under the same transport conditions for a given bias. On the other hand, the volume (thickness) of the absorbing material (low bandgap material) should also be the same or else the total light absorption would not be comparable from one device to the other. In the linear regime applicable to our case, the total light absorption, and therefore the photogenerated current, are directly proportional to the actively absorbing material thickness.[28] Thus, to compare the performance of devices with different amount of absorbing material we have normalized the raw EQE curves by the thickness of low bandgap material in each case. The normalized curves obtained at 0 V and -3 V are shown in **Figure 5**a,b,c,d (see Figure S2a,b,c,d, Supporting Information for the raw curves). I-V measures taken under illumination indicate that the current at -3 V is already saturated in all the analyzed devices (Figure S3, Supporting Information), ensuring complete carrier collection conditions.[29] It can be observed that the bulk sample and the two SLs with 12 nm period provide significant conversion efficiencies in the technologically relevant 1.15 eV spectral region.

Figure 5a shows that there is no difference between the 0 V and -3 V curves for the GaAs device, while the enhancement is only marginal for the GaAsSbN device. We conclude that the carrier collection is almost complete at 0 V in the *bulk* and *GaAs* reference layers as expected for high-quality bulk materials. The remaining voltage dependence is likely an indication of non-radiative recombination at point defects or carrier localization in potential minima induced by strain and composition modulation in the quaternary alloy.[30,31] Moreover, for energies slightly above the bandgap, the EQE is larger in the GaAsSbN bulk layer than in the GaAs reference (both with the same thickness). This enhancement shall be attributed to the increase of the joint density of states, reported for dilute nitrides, arising from a larger electron effective mass and thus a better matching to the hole band dispersion.[32-34]

Regarding the 12 nm period SLs, both the normalized and raw EQE are larger in the *SL-II$_{12}$* sample than in *SL-I$_{12}$* as shown in Figure 5b,c (and Figure S2b,c, Supporting Information), in analogy to the PL emission intensity. This exemplifies the benefits of type-II SLs which, thanks to an improved crystal and interface quality and, in this case, also longer carrier lifetimes, show enhanced optical and transport properties. Yet, the EQE of the *SL-II$_{12}$* and *SL-I$_{12}$* devices also exemplifies the potential drawback of a non-optimized SL structure suffering from slow carrier extraction dynamics. In both samples, EQE increases with the reverse bias limiting severely its application in solar cells. In this case, the carrier extraction at 0 V is not efficient enough because the incomplete formation of minibands for a 12 nm period prevents carrier tunneling through the SL barriers. However, it can be easily restored by reducing the period from 12 to 6 nm as shown in Figure 5d. The fully optimized SL structure (*SL-II$_6$*) shows a normalized EQE which not only is the largest of all samples, but is virtually the same at 0 V as at -3 V. This means complete carrier collection at 0 V and a large overlap of the SL minibands across the structure. A quantum treatment of the transport along the SL[35] is out of the scope of this work but, as a first approximation, we have calculated the transmission coefficient based on analytic expressions using the transfer matrix approach.[36] In such framework, the electronic transmission through a finite array of barriers represents the ability of a confined electron to be transmitted through the periodic structure. **Figure 6** shows the transmission coefficient for electrons below the GaAs bandgap energy as a function of the period thickness (corresponding transmission coefficient for holes are shown in Figure S4, Supporting Information). The results clearly show that electronic coupling and tunneling is strongly enhanced reducing the period from 12 to 6 nm. As a figure of merit, the transmission window halfwidth around the ground energy increases from 0.3 meV to 21 meV in halving the SL period. We conclude that the reduction of the period thickness from 12 to 6 nm leads to a wavefunction overlap and electronic coupling that solve the minority carrier extraction problem without impairing the already mentioned benefits of type-II GaAsSb/GaAsN SL. Although we do not solve the quantum transport equations, our experimental results and calculations for the optimized *SL-II$_6$* sample strongly suggest that vertical transport is more efficient than radiative recombination in this case.

Overall, Figure 5 shows that the normalized EQE at the effective bandgap energy is 72 % higher in the SL structures than in the GaAsSbN bulk sample, indicative of a net advantage coming from all the improvements described above. This is a very promising result, since it could in principle guarantee current matching in GaAs/1.0 eV/Ge solar cells. Indeed, these results suggest that using strain-balanced GaAsSb/GaAsN type-II SLs with thin periods should be a better alternative to thick GaAsSbN layers in this type of solar cells.

## 3. Conclusion

In conclusion, GaAsSb/GaAsN SLs grown on GaAs are shown to overcome the growth problems related to the quaternary GaAsSbN alloy that affect both thick layers and GaAsSbN/GaAs SLs. The spatial separation of both ternaries gives rise to a superior composition and effective bandgap control, as well as to an improved crystal quality and interface abruptness. The type-II band alignment results in long radiative lifetimes that are proven beneficial for carrier extraction. High EQE at 1.15 eV is demonstrated, as well as efficient carrier transport by tunneling though the SL minibands for period thicknesses of ~6 nm. For these thin periods, a net enhancement of EQE over equivalent bulk counterparts is demonstrated under photovoltaic conditions. All these characteristics make GaAsSb/GaAsN SLs an ideal candidate to be monolithically series-connected in ultimately efficient GaAs/1.0 eV/Ge multi-junction solar cells.

**Methods**

*Growth details*: The analyzed samples were all grown by solid source molecular beam epitaxy in a Riber 32 system using GaAs (001) n$^+$ substrates under As$_4$ overpressure conditions. Each sample consists of a 250 nm-thick n-doped GaAs buffer layer, a 200 nm or 750 nm-thick active layer grown at 470 ºC at a growth rate of 1 ML/s, and a 50 nm-thick p-doped GaAs layer deposited on top. The nominal n and p-type doping concentration was $2 \cdot 10^{18}$ cm$^{-3}$. The Sb flux was supplied from a Knudsen effusion cell whereas the atomic N flux came from a radio-frequency plasma source using a 0.1 sccm flow of pure N$_2$. The growth was in-situ monitored by reflective high energy electron diffraction (RHEED) which allowed the verification of a two-dimensional growth mode throughout the whole structure.

*Materials characterization*: TEM was used to obtain dark field 002 images, which were acquired in a JEOL 2100 microscope operating at 200 keV. HR-XRD rocking-curve scans using the Cu-Kα1 line (1.54056 Å) were performed with an X'Pert Pro Pan'alytical commercial system. Low temperature (15 K) PL measurements were carried out using a He-Ne laser. The emitted light was dispersed through a 1 m-spectrometer and detected using a liquid-nitrogen cooled Ge-detector and standard lock-in techniques. TR-PL experiments were performed exciting the sample with 405 nm pulsed laser light. Decay curves were recorded by a time correlated single photon counting system based on a fast-infrared photomultiplier attached to a 0.3 m-focal length spectrometer. The average excitation power density was 0.6 W/cm$^2$ at 10 MHz. Multi-exponential deconvolution analysis was done taking into account the system response measured with a 980 nm ps laser. Time resolution after system response deconvolution is ~200 ps. The optical analysis was completed with PR measurements. PR was performed at room temperature using the 325 nm line of a 15 mW He-Cd laser as pump beam (chopped at 777 Hz), the monochromatic (1/8 m-spectrometer) probe beam from a 150 W QTH lamp, and a cooled InGaAs-photodetector.

*Electronic band structure calculation*: The calculation is based on finite differences methods. Single band effective mass approximation was used with input parameters for the Sb and N contents obtained from the HR-XRD spectra of the ternary samples, and the period thickness estimated from the TEM measurements of each sample. The bandgap energy and band offsets of GaAsN were obtained considering the BAC predictions in first order perturbation theory,[10] with the specific values for the electron effective mass and BAC parameters described elsewhere.[37] Regarding the GaAsSb layers, their bandgap energy and band offsets were estimated using experimental results for GaAsSb pseudomorphically grown on GaAs.[38] The hole effective mass was obtained from a linear interpolation between the binaries.

*Device fabrication*: The thickest samples (750 nm active layer) were processed in 200 μm-diameter mesa-etched devices using standard fabrication techniques. The mesa structures were defined by wet etching using a $H_3PO_4$-$H_2O_2$-$H_2O$ (1:1:8) solution. The p-type contact, deposited on top of the mesa, was Au/Au-Zn/Au (100/800/2000 Å). A common n-type contact consisting on Au-Ge/Au (800/2000 Å) was evaporated on the substrate side. The contacts were exposed to an annealing process at 400 °C during 1 minute.

*Device characterization*: Photocurrent measurements were carried out using light from a QTH lamp which was dispersed through a 0.34 m monochromator and directed through the optical path to the sample. A K230 Keithley sourcemeter as well as a K617 Keithley electrometer were employed. To obtain the EQE, the photocurrent data was first converted into responsivity dividing by the power per unit area of the QTH lamp (measured using a calibrated Si photodiode and a pyroelectric photodetector) multiplied by the diode top metal-free area. Finally, responsivity was converted to EQE multiplying by the photon energy divided by the electron charge.

*Transmission coefficients calculation*: The parameters used in the calculation for the bandgap energies, effective masses and band offsets are those used in the electronic band structure calculations. Calculations were done separately for electrons and holes transmitted through 12 consecutive barriers at flat band condition.

**Acknowledgements**

We acknowledge funding from the Spanish MINECO through projects MAT2013-47102-C2-2-R and MAT2016-77491-C2-1-R and from CSIC through project 2015-50-I021. DFM acknowledges financial support of Comunidad de Madrid (S2013/MAE-2780). We would also like to thank Maika Sabido (ISOM-UPM) for processing the samples.

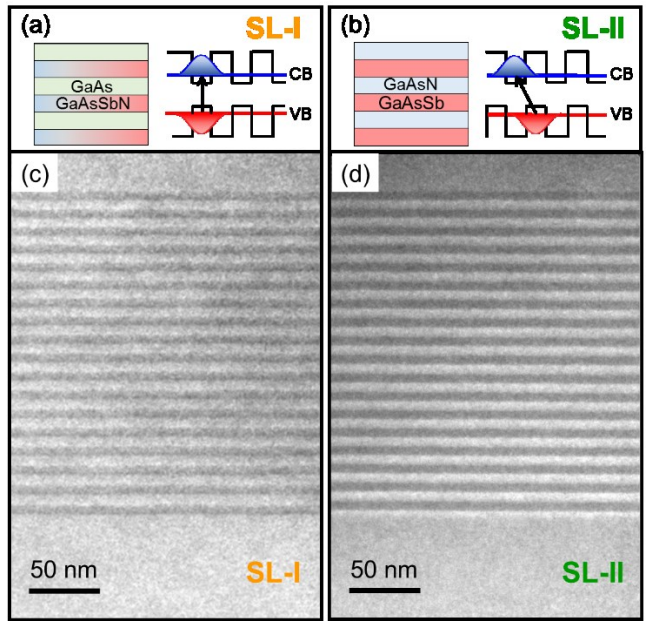

**Figure 1.** Sketch of the epitaxial and band structure of samples a) *SL-I* and b) *SL-II*. Dark field 002 TEM images from sample c) *SL-I* and d) *SL-II*.

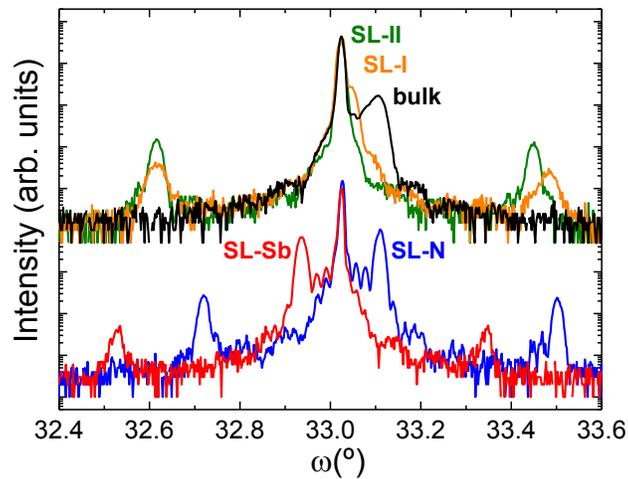

**Figure 2.** Omega-2 theta HR-XRD scans around the (004) GaAs Bragg reflection performed on the ternary samples *SL-Sb* and *SL-N* (below) and on the samples *SL-I, SL-II*, and *bulk* (above). Only the GaAsSb/GaAsN structure (*SL-II*) fits perfectly the lattice matching condition expected from the combination of both ternaries.

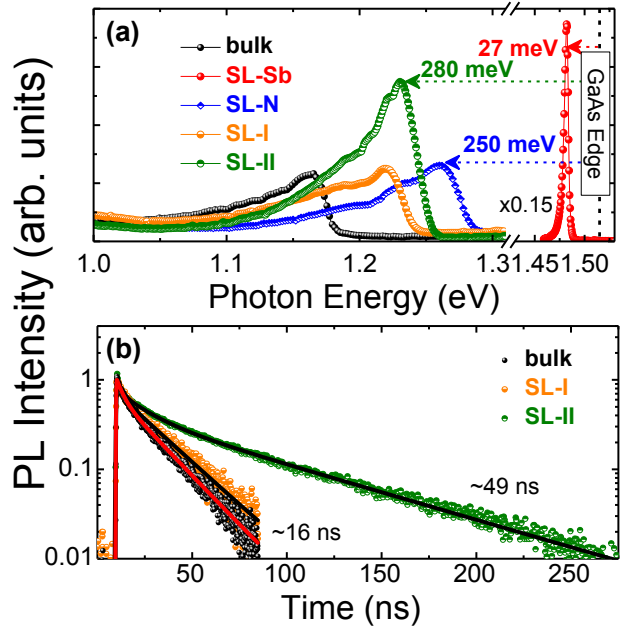

**Figure 3.** a) 15 K PL spectra from the first set of samples. b) Time-resolved PL decays measured at the PL peak energy for samples *bulk*, *SL-I* and *SL-II*. The deconvoluted decay time parameters are presented in Table S1, Supporting Information.

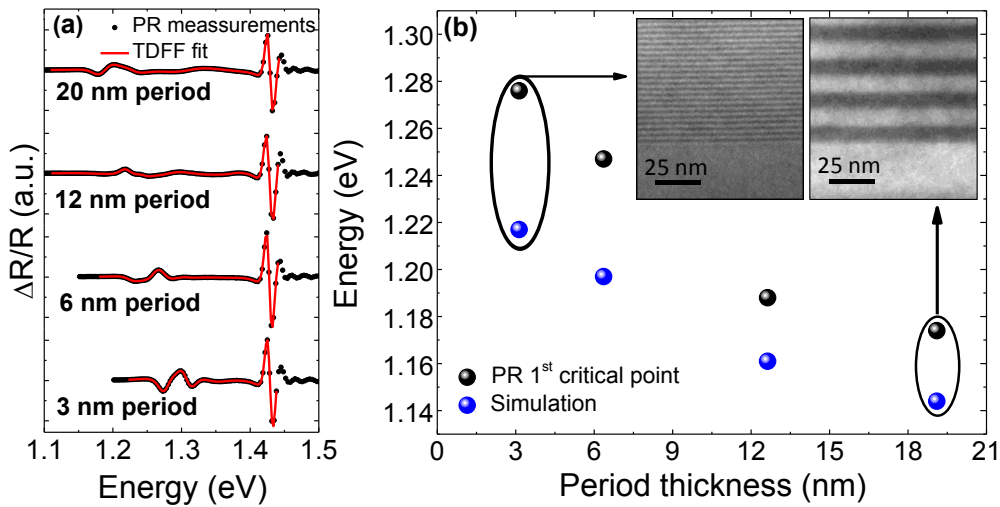

**Figure 4.** a) PR experimental spectra (black dots) and TDFF fittings (red lines) for the SL structures with different periods and b) effective bandgap energies at room temperature as a function of the period thickness obtained from PR measurements (black dots) and from simulations based on finite differences method (blue dots). The inset shows dark field 002 TEM images of the samples with the thinnest and thickest period, showing an accurate control of periodicity even for very thin periods of 3 nm (5 MLs GaAsSb/5 MLs GaAsN).

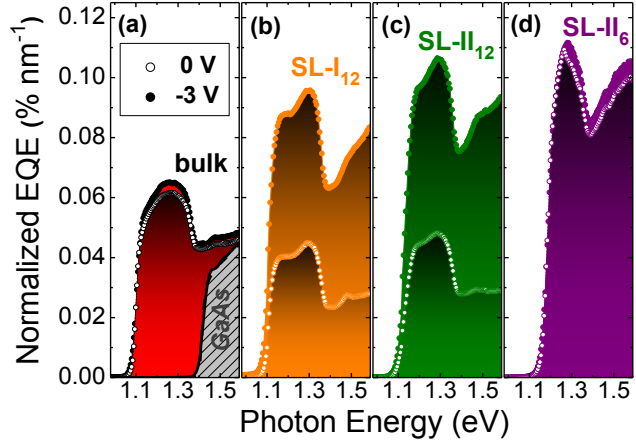

**Figure 5.** Normalized EQE spectra at room temperature at 0 V (empty dots) and -3 V (filled dots) from samples a) *bulk* together with sample *GaAs* as a reference b) *SL-I$_{12}$* c) *SL-II$_{12}$* and d) *SL-II$_6$*.

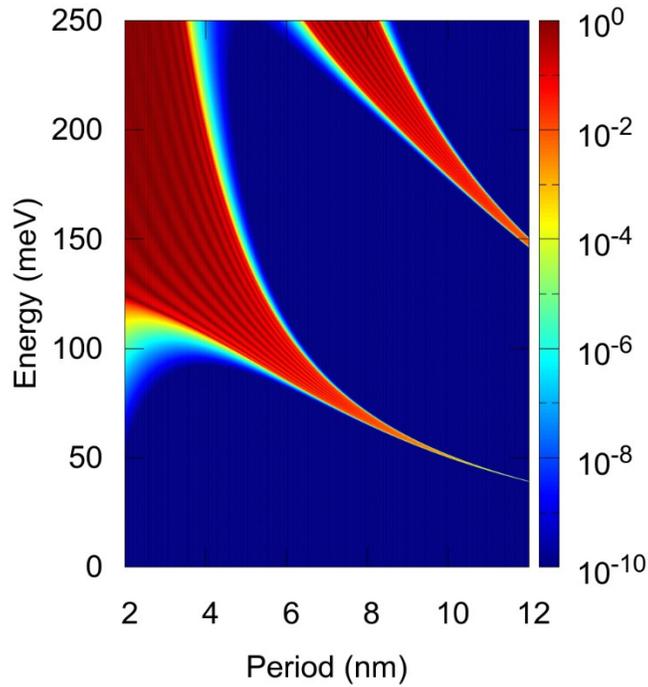

**Figure 6.** Contour plot of the electron transmission coefficient for a 12 barrier/quantum well array as a function of period thickness. The height of the barrier is 250 meV and the effective mass 0.146 m$_0$, which are the parameters used to model the *SL-II* sample in Figure 4b. The color bar is in log scale.

**Supporting Information**

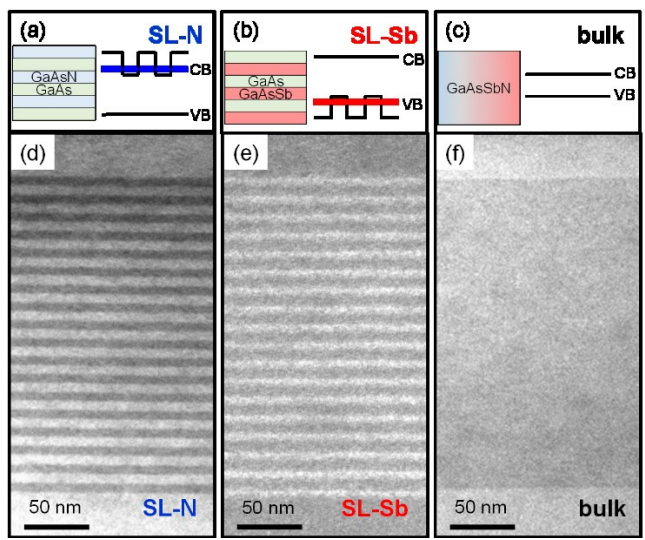

**Figure S1.** Sketch of the epitaxial and band structure of samples a) *SL-N* b) *SL-Sb* and c) *bulk*. Dark field 002 TEM images from sample d) *SL-N* e) *SL-Sb* and f) *bulk*.

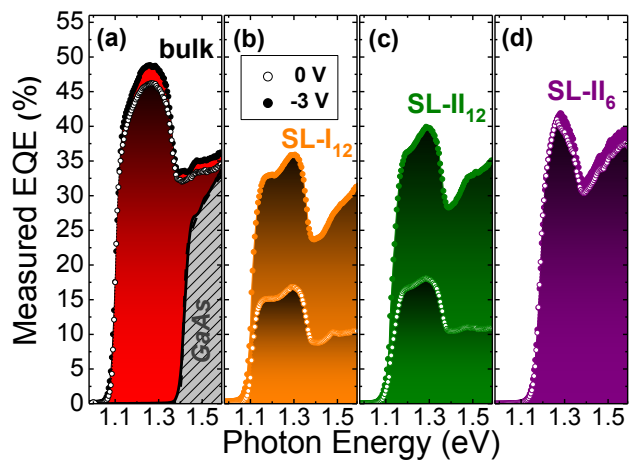

**Figure S2**. Raw EQE spectra at room temperature at 0 V (empty dots) and -3 V (filled dots) from samples a) *bulk* together with sample *GaAs* as reference b) *SL-I$_{12}$* c) *SL-II$_{12}$* and d) *SL-II$_6$*.

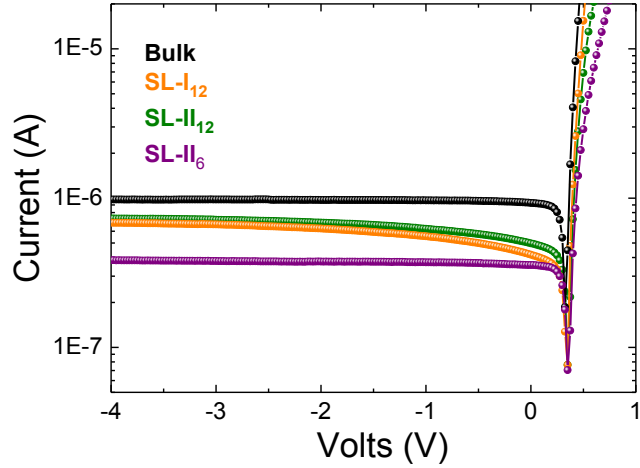

**Figure S3**. I-V curves of the samples *bulk*, *SL-I$_{12}$*, *SL-II$_{12}$* and *SL-II$_6$* under 1.2 eV monochromatic illumination at room temperature. At -3 V current has already saturated in all cases, so this voltage can be considered to provide complete carrier collection condition.

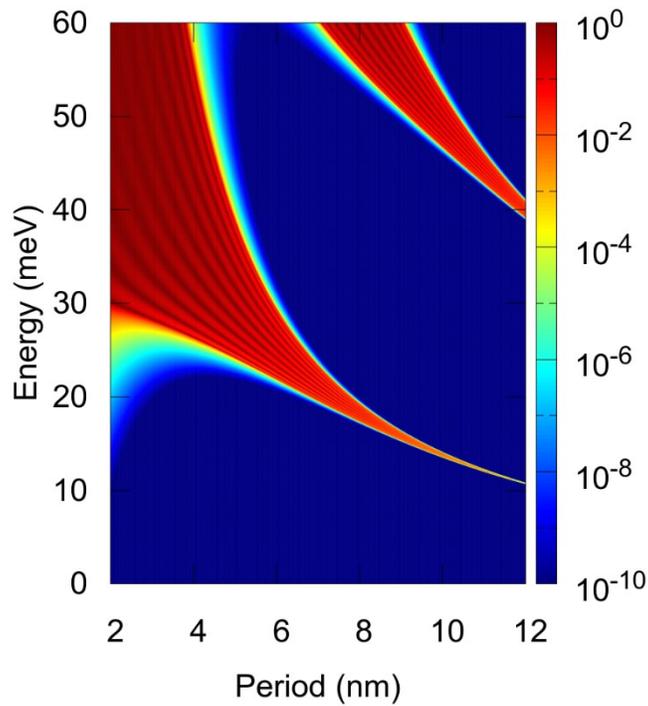

**Figure S4**. Contour plot of the holes transmission coefficient for a 12 barrier/quantum well array as a function of period thickness. The height of the barrier is 70 meV and the effective mass 0.506 m$_0$, which are the parameters used to model the *SL-II* sample in Figure 4b. The color bar is in log scale.